\documentclass[aip,graphicx,amsmath,amssymb,english,reprint]{revtex4-2}
\usepackage[T1]{fontenc}
\usepackage[utf8]{inputenc}
\usepackage[margin=0.6in]{geometry}
\usepackage{setspace}
\usepackage{babel}
\usepackage{amssymb}
\usepackage{leftidx}
\usepackage[titletoc,toc,title]{appendix}
\usepackage{braket}
\usepackage{bm}
\usepackage{xcolor}
\usepackage{comment}
\usepackage{graphicx}
\begin{document}

\title{Parametrically driven THz magnon-pairs: predictions towards ultimately fast and minimally dissipative switching}

\author{G. Fabiani} \email{gfabiani@science.ru.nl}
\author{J. H. Mentink}
\affiliation{Radboud University, Institute for Molecules and Materials (IMM) Heyendaalseweg 135, 6525 AJ Nijmegen, The Netherlands}

\date{\today}

\begin{abstract}
Findings ways to achieve switching between magnetic states at the fastest possible time scale that simultaneously dissipates the least amount of energy is one of the main challenges in magnetism. 
Antiferromagnets exhibit intrinsic dynamics in the THz regime, the highest among all magnets and are therefore ideal candidates to address this energy-time dilemma. Here we study theoretically THz-driven parametric excitation of antiferromagnetic magnon-pairs at the edge of the Brillouin zone and explore the potential for switching between two stable oscillation states. Using a semi-classical theory, we predict that switching can occur at the femtosecond time scale with an energy dissipation down to a few zepto Joule. This result touches the thermodynamical bound of the Landauer principle, and approaches the quantum speed limit up to 5 orders of magnitude closer than demonstrated with magnetic systems so far.
\end{abstract}


\maketitle 

\section{Introduction}
The explosive growth of digital data is fuelling the quest for faster and more energy-efficient ways to process and store digital information in magnetic storage devices.
Following the discovery of all-optical switching (AOS) in ferrimagnetic GdFeCo\cite{Stanciu2007}, which was subsequently demonstrated in related and other ferrimagnetic alloys \cite{Liu2015, Banarjee2020, Davies2020, Banerjee2021}, ultrafast manipulation of magnetic order has emerged into a cutting edge approach towards both faster and more energy efficient switching. Despite that the mechanism relies on ultrafast heating \cite{Radu2011,Ostler2012,Mentink2012}, in GdFeCo a write-read cycle was performed lasting only about 30 ps, with an unprecedented energy dissipation of ~16 fJ when scaled to nanometer dimensions \cite{Vahaplar2009}. Interestingly, the energy and time attained are several orders of magnitude smaller than existing technologies such as hard disk drives\cite{Hylick2008} and spin-torque based memories\cite{Wang2013}. AOS was later observed in metallic ferromagnets as well \cite{Lambert2014, ElHadri2016, Medapalli2017} and more recently also in the transparent dielectric YIG:Co. Distinct from metals, in the latter case, the magnetic recording proceeds non-thermally \cite{Stupakiewicz2017,Davies2019,Stupakiewicz2021}, allowing for a write-read event lasting less then 20 ps, with an estimated energy dissipation down to only 22 aJ\cite{Stupakiewicz2017}. 

Antiferromagnets (AFMs) are a class of materials where even faster and more energy efficient switching can be expected. These materials have attracted considerable interest in the last decades\cite{Kimel2004, Kimel2009, Zhang2014, Jungwirth2016} as they offer several advantages, such as robustness to magnetic disturbance and the absence of stray fields. Importantly, the frequency of their collective spin excitations is enhanced by the exchange interaction between the magnetic sublattices, boosting it to the THz range \cite{Kittel1951,Kampfrath2011}. 
Switching in AFMs has been discussed in several works, featuring mechanisms based on excitation with THz-pulses\cite{Wienholdt2012,Schlauderer2019}, femtosecond laser pulses \cite{Dannegger2021} and in spintronics via spin-torque transfer mechanisms\cite{Cheng2015, Moriyama2018, Chirac2020} and electric currents\cite{ Wadley2016, Olejnik2017}. 
Nevertheless, the switching pathways involve coherent excitation of long wavelength magnons, reaching time scales comparable to that in ferrimagnets.

In principle, even faster dynamics in AFMs can be launched by exciting magnons at the edge of the Brillouin zone. These have a frequency solely determined by the exchange interactions and can be as high as few tens of THz, having at the same time wavelength down to the sub-nanometer range. Recently, femtosecond dynamics of pairs of these high-energy magnons have been observed and addressed theoretically \cite{Zhao2004, Bossini2016, Bossini2019, Fabiani2021, Fabiani2021-2}, revealing unique features such as high entanglement between magnons\cite{Zhao2004, Fabiani2021-2} and coherent control of longitudinal oscillations of the AFM vector\cite{Bossini2016, Bossini2019}. However, ways to employ such magnons to achieve switching at the femtosecond timescale have, to the best of our knowledge, not been discussed so far.

Here, we theoretically show that periodic modulations of the exchange interactions can trigger a parametric resonance of zone-edge magnon pairs in AFMs. By developing a semi-classical description of the magnon-pair dynamics, we show that it is possible to steer the system between two opposite phases of the parametric oscillations. Distinct from switching between thermodynamically stable states, we consider an analog of the parametron architecture introduced in the '50 \cite{Goto1959} where parametric oscillators are exploited to perform logic operations. We demonstrate a simple instance of switching in a magnon-based parametron that employs magnon pairs at the edge of the Brillouin zone.
The switching between the two opposite phases of the oscillations can be attained in the femtosecond timescale and with energy dissipation down to $\sim 3$ zJ. This is both more energy-efficient and much faster than previously proposed switching mechanisms and touches the physical limits of computing, namely the thermodynamical Landauer's limit \cite{Landauer1961} and the quantum speed limit \cite{Deffner2017}.

This work is structured as follows: 
In Section \ref{sec:2} we review the magnon-pair dynamics within the linear spin wave theory (LSWT) and we study parametric driving of magnon pairs. In Section \ref{sec:3} we develop a semi-classical approach which allows us to take into account magnon damping on a phenomenological level. This will be employed in Section \ref{sec:4} to develop the concept of magnon-based parametron. Results for the switching time and energy are given in Section \ref{sec:5}, where a discussion about the fundamental limits of binary computing is presented. Section \ref{sec:6} concludes with a summary and outlook for future research in this direction.

\section{Parametric excitation of magnon-pair dynamics}\label{sec:2}
In order to investigate the parametric resonance of magnon pairs, we consider the minimal model for quantum dynamics in antiferromagnets, namely the antiferromagnetic Heisenberg model in a square lattice, defined by
\begin{equation}\label{eq:1}
\hat{\mathcal{H}}_0 = J_{\textrm{ex}} \sum_{\langle ij\rangle } \hat{S}_i \cdot \hat{S}_{j},
\end{equation}
where $\hat{S}_i=\hat{S}(\mathbf{r}_i)$ are spin-$1/2$ operators, with $\mathbf{r}_i=(x_i,y_i)$, $J_{\text{ex}}$ is the exchange interaction ($J_{\textrm{ex}}>0$) and $\langle \cdot \rangle$ restricts the sum to nearest neighbours. Recent studies on this model have shown that impulsive \cite{Zhao2004, Bossini2019} and quench-like \cite{Fabiani2021,Fabiani2021-2} perturbations of the exchange interaction lead to coherent oscillations of magnon pairs. Here we focus on parametric oscillations which can be excited by periodic modulation of exchange interactions by THz field transients \cite{Mentink2015, Mentink2017, Eckstein2017} and optical pulses \cite{Zhao2004, Bossini2016, Bossini2019} described by
\begin{equation}\label{eq:2}
\delta \hat{\mathcal{H}}(t) = \Delta J_{\text{ex}}f(t)\sum_{i,\pmb{\delta}}\big(\mathbf{e}\cdot \pmb{\delta}\,\big)^2\hat{S}(\mathbf{r}_i) \cdot \hat{S}(\mathbf{r}_i+\pmb{\delta}),
\end{equation}
where $\mathbf{e}$ is a unit vector that determines the polarization of the electric field of the light pulse which causes the perturbation, $\pmb{\delta}$ connects nearest neighbour spins, $\Delta J_{\text{ex}}/J_{\text{ex}} \ll 1$ and $f(t) = \cos \omega_0 t$ models the pulse shape. In the following we will set $\mathbf{e}$ along the y-direction of the lattice and the lattice constant $a=1$.

To treat Eqs. \eqref{eq:1}-\eqref{eq:2}, we consider the simplest approximation based on a linear spin wave theory (LSWT) expansion of the Hamiltonian $\hat{\mathcal{H}}=\hat{\mathcal{H}}_0 +\delta\hat{\mathcal{H}}$ in terms of magnon operators. This proceeds by expanding the spin operators in terms of Holstein-Primakoff (HP) bosons keeping only the linear terms of the expansion. The resulting Hamiltonian is then diagonalized via a Bogoliubov transformation in reciprocal space, yielding 
\begin{eqnarray}\label{eq:H_bog}
\hat{\mathcal{H}} = \sum_{\mathbf{k}}  \Big[\big(\omega_\mathbf{k} + f(t)\delta\omega_\mathbf{k}\big)2\hat{K}^z_\mathbf{k} + 
f(t) V_\mathbf{k}\big(\hat{K}^+_\mathbf{k} + \hat{K}^-_\mathbf{k}\big)\Big],
\end{eqnarray}
with
\begin{eqnarray}\label{eq:Kz_def}
\hat{K}^z_\mathbf{k}=\frac{1}{2}(\hat{\alpha}^\dagger_\mathbf{k}\hat{\alpha}_\mathbf{k} + \hat{\beta}_{-\mathbf{k}}\hat{\beta}^\dagger_{-\mathbf{k}}), \\ \label{eq:Kpm_def}
\hat{K}^+_\mathbf{k}=\hat{\alpha}^\dagger_\mathbf{k}\hat{\beta}^\dagger_{-\mathbf{k}},\quad \hat{K}^-_\mathbf{k}=\hat{\alpha}_\mathbf{k}\hat{\beta}_{-\mathbf{k}},
\end{eqnarray}
where $\hat{\alpha}_\mathbf{k}$, $\hat{\beta}_\mathbf{k}$ ($\hat{\alpha}_\mathbf{k}^\dagger$, $\hat{\beta}_\mathbf{k}^\dagger$) are the magnon annihilation (creation) operators with wave vector $\mathbf{k}$ (note that there are two species of bosons due to the bipartite nature of the model considered) \cite{Bossini2019}. Here $\omega_\mathbf{k}=zSJ_\text{ex}\sqrt{1-\gamma_\mathbf{k}^2}$ is the single-magnon dispersion, where $\gamma=\frac{1}{z}\sum_{\pmb{\delta}}e^{i\mathbf{k}\cdot \pmb{\delta}}$ and $z=4$ is the coordination number. The second term in Eq. \eqref{eq:H_bog} yields a time-dependent renormalization of the magnon dispersion, while the third term  is responsible for the creation and annihilation of pairs of counter-propagating magnons, with coefficients respectively
\begin{eqnarray}
\delta \omega_\mathbf{k}=2zS\Delta J_\text{ex}\frac{1-\xi_\mathbf{k}\gamma_\mathbf{k}}{\sqrt{1-\gamma^2_\mathbf{k}}},\quad V_\mathbf{k} =2zS \Delta J_\text{ex}\frac{\gamma_\mathbf{k}-\xi_\mathbf{k}}{\sqrt{1-\gamma^2_\mathbf{k}}},
\end{eqnarray}
with $\xi_\mathbf{k}=\frac{1}{2}\sum_{\pmb{\delta}}(\mathbf{e}\cdot \pmb\delta)^2e^{i\mathbf{k}\cdot\pmb{\delta}}$.
The equations of motion can be obtained from the Heisenberg equations for the magnon-pair operators defined in Eq. \eqref{eq:Kz_def}-\eqref{eq:Kpm_def}, that yield
\begin{eqnarray}\label{eq:Kz_eom}
\frac{d\braket{\hat{K}^z_{\mathbf{k}}(t)}}{dt}&=& if(t) V_{\mathbf{k}}(\braket{\hat{K}^-_{\mathbf{k}}}-\braket{\hat{K}^+_{\mathbf{k}}}),\\ \label{eq:Kpm_eom}
\frac{d\braket{\hat{K}^\pm_{\mathbf{k}}(t)}}{dt}&=& \pm 2i\big(\omega_{\mathbf{k}}+f(t)\delta\omega_{\mathbf{k}}\big)\braket{\hat{K}^{\pm}_{\mathbf{k}}} \pm 2if(t) V_{\mathbf{k}}\braket{\hat{K}^z_{\mathbf{k}}}.
\end{eqnarray}
For the ground state of $\hat{\mathcal{H}}_0$ we have  $\braket{\hat{K}_\mathbf{k}^z(0)}=1/2$ and $\braket{\hat{K}_\mathbf{k}^{\pm}(0)}=0$, which we take as initial conditions in the remainder. To get insight into the parametric resonance of magnon-pair dynamics, we consider two approximations that allow to solve these equations analytically: (i) we neglect the term proportional to $\delta \omega_\mathbf{k}$ in Eq. \eqref{eq:H_bog} which amounts to a small renormalization of the oscillation frequency; (ii) we approximate 
$\cos(\omega_0 t) (\hat{K}^+_{\mathbf{k}}+\hat{K}^-_{\mathbf{k}})\simeq \frac{1}{2}\big(e^{-i\omega_0 t}\hat{K}^+_{\mathbf{k}} + e^{i\omega_0 t}\hat{K}^-_{\mathbf{k}}\big)$
which is exact up to a term proportional to $\hat{K}^+_{\mathbf{k}}-\hat{K}^-_{\mathbf{k}}$. The rationale behind these approximations is that now we can easily switch to a rotating frame where the Hamiltonian becomes time-independent, by means of the following transformations
\begin{equation}
\tilde{K}^z_{\mathbf{k}} = \hat{K}^z_{\mathbf{k}} , \quad \tilde{K}^+_{\mathbf{k}} = e^{-i\omega_0 t}\hat{K}^+_{\mathbf{k}}, \quad \tilde{K}^-_{\mathbf{k}} = e^{i\omega_0 t}\hat{K}^-_{\mathbf{k}}.  
\end{equation}
The Heisenberg equations for the transformed operators in the rotating frame read
\begin{eqnarray}
\frac{d\braket{\tilde{K}^z_{\mathbf{k}}(t)}}{dt}&=& i \tilde{V}_{\mathbf{k}}(\braket{\tilde{K}^-_{\mathbf{k}}}-\braket{\tilde{K}^+_{\mathbf{k}}})\\ 
\frac{d\braket{\tilde{K}^\pm_{\mathbf{k}}(t)}}{dt}&=& \pm 2i\tilde{\omega}_\mathbf{k}\braket{\tilde{K}^{\pm}_{\mathbf{k}}} \pm 2i\tilde{V}_{\mathbf{k}}\braket{\tilde{K}^z_{\mathbf{k}}},
\end{eqnarray}
where $\tilde{\omega}_\mathbf{k}=\omega_{\mathbf{k}}-\frac{\omega_0}{2}$ and $\tilde{V}_\mathbf{k}=V_\mathbf{k}/2$, with the solutions
\begin{eqnarray}\label{eq:rot_frame_sol}
\left \{ \begin{array}{rl}
\braket{\tilde{K}_\mathbf{k}^z} &= \frac{2\tilde{V}_\mathbf{k}^2}{a_\mathbf{k}^2}(\cosh(a_\mathbf{k} t)-1)+\frac{1}{2};\\
\braket{\tilde{K}_\mathbf{k}^+ + \tilde{K}_\mathbf{k}^-} &= -\frac{4\tilde{V}_\mathbf{k}\tilde{\omega}_\mathbf{k}}{a_\mathbf{k}^2}(\cosh(a_\mathbf{k}t)-1),
\end{array}
\right.
\end{eqnarray}
where $a_{\mathbf{k}}^2 = 4(\tilde{V}_\mathbf{k}^2 - \tilde{\omega}_{\mathbf{k}}^2)$.  Parametric oscillations with exponentially increasing amplitude are excited when $a^2_\mathbf{k}>0$.  This is always realized at the resonant condition $\omega_0 = 2\omega_\mathbf{k}$ and shares close similarities with previous studies showing that Eq. \eqref{eq:H_bog} is the Hamiltonian of a degenerate parametric oscillator \cite{Gerry1982, Gerry1985}. As we will see in the next section, the existence of diverging solutions is a consequence of the Lie algebra of the magnon-pair operators that confines the orbit of the solutions to the (unbounded) hyperbolic plane. In principle, the divergences can be cured by including magnon-magnon interactions beyond LSWT \cite{Kloss2010}. 
Here, however, we retain the linear approximation and to ensure consistency we will keep the perturbation small enough such that the number of HP boson excited $\braket{n_{\text{HP}}}$ at each time remains close to the range of applicability of LSWT \cite{Manousakis1991}
\begin{equation}
   \frac{\braket{n_{\text{HP}}}}{2S} = -\frac{1}{2S} + \frac{1}{SN}\sum_\mathbf{k}\Big[\frac{ \braket{\hat{K}^z_\mathbf{k}} -\frac{\gamma_\mathbf{k}}{2}\big(\braket{\hat{K}^+_\mathbf{k}}+\braket{\hat{K}^-_\mathbf{k}}\big) }{\sqrt{1-\gamma^2_\mathbf{k}}}\Big] \ll 1.
\end{equation}

Importantly, in the linear approximation each magnon-pair mode stays coherent and decay due to coupling with other degrees of freedom such as phonons is absent. In the following, we derive a semi-classical description of the magnon-pair dynamics, where the coupling with an external bath is introduced phenomenologically via a Rayleigh dissipation function. In this way, dissipation can be taken into account in our model while keeping the simple structure of the equations of motion.

\section{Phenomenological theory of magnon-pair dynamics}\label{sec:3}
The magnon-pair operators defined in Eqs. \eqref{eq:Kz_def}-\eqref{eq:Kpm_def} are Perelomov operators \cite{Perelomov1975} corresponding to the generators of the SU(1,1) Lie algebra \cite{Gerry1982}. These operators were introduced in the context of SU(1,1) coherent states and in this respect Eq. \eqref{eq:H_bog} is the generic form of a SU(1,1) coherent-state preserving Hamiltonian. 
By exploiting the properties of the SU(1,1) Lie algebra we can construct a semi-classical description of magnon pairs in analogy with the SU(2) spin case of classical magnetization dynamics.
Since the Hamiltonian Eq. \eqref{eq:H_bog} is quadratic, the magnon-pair physics within LSWT is fully determined by the expectation values of the magnon-pair operators. This allows us to define a semi-classical approach where quantum operators are replaced with (classical) expectation values. To this end, we define the pseudospin operator 
$\pmb{\mathcal{K}}_\mathbf{k} = \big({\mathcal{K}_\mathbf{k}^x},{\mathcal{K}_\mathbf{k}^y},{\mathcal{K}_\mathbf{k}^z}\big)$, where
\begin{eqnarray}
{\mathcal{K}_\mathbf{k}^x} &=& \frac{1}{2}\braket{\hat{K}_\mathbf{k}^+ +\hat{K}_\mathbf{k}^-},\; {\mathcal{K}_\mathbf{k}^y}=\frac{1}{2i}\braket{\hat{K}_\mathbf{k}^+ -\hat{K}_\mathbf{k}^-},\; \mathcal{K}_\mathbf{k}^z=\braket{\hat{K}_\mathbf{k}^z}. \nonumber
\end{eqnarray} 
Similarly, we can define a semi-classical Hamiltonian as
\begin{equation}
H_\text{cl}= \braket{\hat{\mathcal{H}}} = \sum_\mathbf{k}2  \Big[\big(\omega_\mathbf{k} + f(t)\delta\omega_\mathbf{k}\big)\mathcal{K}^z_\mathbf{k} + f(t) V_\mathbf{k}\mathcal{K}^x\Big].
\end{equation}
In terms of the pseudospin operator $\pmb{\mathcal{K}}_\mathbf{k}$, the equations of motion  \eqref{eq:Kz_eom}-\eqref{eq:Kpm_eom} can be recast in the form of a Bloch equation
\begin{equation}\label{eq:LL}
\frac{d\pmb{\mathcal{K}}_\mathbf{k}}{dt} =  \pmb{\mathcal{K}}_\mathbf{k}\times \mathbf{B}_\mathbf{k} , 
\end{equation}
with 
\begin{eqnarray}
\mathbf{B}_\mathbf{k} &=& \frac{\partial H_\text{cl}}{\partial \pmb{\mathcal{K}}_\mathbf{k}}=\bigg(\frac{\partial H_\text{cl}}{\partial {\mathcal{K}_\mathbf{k}^x}},\frac{\partial H_\text{cl}}{\partial {\mathcal{K}_\mathbf{k}^y}},-\frac{\partial H_\text{cl}}{\partial {\mathcal{K}^z_\mathbf{k}}}  \bigg) \nonumber \\ &=& 2\big(V_\mathbf{k}f(t),\,0,\,-\omega_\mathbf{k}-\delta\omega_\mathbf{k}f(t)\big),
\end{eqnarray}
where  the cross product is 
\begin{eqnarray}
(\mathbf{a}\times \mathbf{b})^x &=& a^yb^z-a^zb^y, \quad (\mathbf{a}\times \mathbf{b})^y = a^zb^x-a^xb^z, \nonumber \\
(\mathbf{a}\times \mathbf{b})^z &=& -(a^xb^y-a^yb^x).
\end{eqnarray}
Note that both the gradient and the cross product have a minus sign in the third component, as they are defined in the hyperbolic SU(1,1) plane. In the same way, a scalar product can be defined as
\begin{equation}
\mathbf{a}\cdot \mathbf{b} = a^x b^x + a^y b^y -a^z b^z.
\end{equation}
Eq. \eqref{eq:LL} is the analogous of the SU(2) Bloch equation  with effective field $\mathbf{B}_\mathbf{k}$. 
As such, it describes precessional motion of the pseudospin $\pmb{\mathcal{K}}_\mathbf{k}$, which however, is not confined in the unit sphere but it takes place in the hyperbolic SU(1,1) "sphere". 
This can be seen by taking the scalar product of both sides of Eq. \eqref{eq:LL} with $\pmb{\mathcal{K}}_\mathbf{k}$, from which follows that, analogously to the spin invariant $\pmb{{S}}^2$ of the SU(2) algebra,  $\pmb{\mathcal{K}}_\mathbf{k} \cdot \pmb{\mathcal{K}}_\mathbf{k}$ is a constant of motion and therefore the orbits live in the hyperboloid defined by $(\mathcal{K}_\mathbf{k}^x)^2 + (\mathcal{K}_\mathbf{k}^y)^2- (\mathcal{K}_\mathbf{k}^z)^2=-1/4$. The factor $-1/4$ comes from evaluating $\pmb{\mathcal{K}}_\mathbf{k} \cdot \pmb{\mathcal{K}}_\mathbf{k}$ in the ground state at $t=0$, the negative sign relates to the negative curvature of the hyperbolic "sphere".
Moreover, note that  $\pmb{\mathcal{K}}_\mathbf{k} \cdot \pmb{\mathcal{K}}_\mathbf{k}$ is the classical analogue of the Casimir invariant of the SU(1,1) algebra
\begin{equation}
 \hat{Q}_\mathbf{k}=-(\hat{K}^z_\mathbf{k})^2+(\hat{K}^+_\mathbf{k}\hat{K}^-_\mathbf{k}+ \hat{K}^-_\mathbf{k}\hat{K}^+_\mathbf{k})/2. 
\end{equation}
Similar to the SU(2) case, we can parametrize the pseudospin operator in terms of angular variables living in the SU(1,1) sphere as
\begin{eqnarray}
\mathcal{K}_\mathbf{k}^x &=& \mathcal{K} \sinh \theta_\mathbf{k} \cos \phi_\mathbf{k},\quad \mathcal{K}_\mathbf{k}^y = \mathcal{K}\sinh \theta_\mathbf{k} \sin \phi_\mathbf{k},\nonumber\\
\mathcal{K}_\mathbf{k}^z &=& \mathcal{K} \cosh \theta_\mathbf{k}, 
\end{eqnarray}
with $-\infty <\theta_\mathbf{k} < \infty$, $0\leq \phi_\mathbf{k} \leq2\pi$ and $\mathcal{K} = \sqrt{-\pmb{\mathcal{K}}_\mathbf{k} \cdot \pmb{\mathcal{K}}_\mathbf{k}}=1/2$. In this way we can easily define a Lagrangian theory for the magnon-pair dynamics in terms of $\theta_\mathbf{k}$ and $\phi_\mathbf{k}$ as done for semi-classical spin dynamics\cite{Zvezdin1979}. 
In order to derive a Lagrangian, what we demand is that the Euler-Lagrange (E-L) equations for $\theta_\mathbf{k}$ and $\phi_\mathbf{k}$ will match the equations for the magnon-pair operators  Eqs. \eqref{eq:Kz_eom}-\eqref{eq:Kpm_eom} or equivalently Eq. \eqref{eq:LL}. This can be obtained with the following Lagrangian
\begin{eqnarray}
\mathcal{L} = \sum_\mathbf{k}{\mathcal{K}}\big[ \dot{\phi}_\mathbf{k}(\cosh \theta_\mathbf{k}-1)\big] - H_\text{cl}.
\end{eqnarray}
This semi-classical approach allows us to describe the coupling with an environment via a Rayleigh dissipation function. Generalizing the SU(2) case of the classical spin dynamics \cite{Zvezdin1979} to the hyperbolic variables $\theta_\mathbf{k}$ and $\phi_\mathbf{k}$, we can define the Rayleigh dissipation function
\begin{equation}\label{eq:Rayleigh}
F[\phi,\theta,\dot{\phi},\dot{\theta}]= \frac{\eta}{2}\sum_\mathbf{k}\mathcal{K} \Big[\sinh^2(\theta_\mathbf{k})\,\dot{\phi}^2_\mathbf{k} + \dot{\theta}^2_\mathbf{k}\Big],
\end{equation}
where $\eta$ is the phenomenological damping coefficient. 
The equations of motion are then derived from the E-L equations
\begin{eqnarray}
\frac{d}{dt}\frac{\partial \mathcal{L}}{\partial \dot{\theta}_{\mathbf{k}}} - \frac{\partial \mathcal{L}}{\partial\theta_{\mathbf{k}}} = -\frac{\partial F}{\partial \dot{\theta}_\mathbf{k}},\\
\frac{d}{dt}\frac{\partial \mathcal{L}}{\partial \dot{\phi}_{\mathbf{k}}} - \frac{\partial \mathcal{L}}{\partial\phi_{\mathbf{k}}} = -\frac{\partial F}{\partial \dot{\phi}_\mathbf{k}},
\end{eqnarray} 
and read
\begin{eqnarray}\label{eq:theta}
\dot{\theta}_\mathbf{k} &=& 2 V_\mathbf{k}f(t)\sin\phi_\mathbf{k} - \eta\sinh(\theta_\mathbf{k})\dot{\phi}_\mathbf{k}, \\ \label{eq:phi}
\dot{\phi}_\mathbf{k} &=& 2(\omega_\mathbf{k} + f(t)\delta\omega_\mathbf{k}) \nonumber\\
&+& 2 V_\mathbf{k}f(t) \cos\phi_\mathbf{k}\coth \theta_\mathbf{k}+\frac{\eta}{\sinh\theta_\mathbf{k}}\dot{\theta}_\mathbf{k}.
\end{eqnarray}
Note that the effect of damping could be introduced phenomenologically in Eq. \eqref{eq:LL} in the form of a Gilbert damping term\cite{Gilbert2004}. In particular, the choice of the Rayleigh dissipation function Eq. \eqref{eq:Rayleigh} corresponds to 
\begin{equation}
\frac{d\pmb{\mathcal{K}}_\mathbf{k}}{dt} = \pmb{\mathcal{K}}_\mathbf{k}\times \mathbf{B}_\mathbf{k} + \frac{\eta} {\mathcal{K}} \pmb{\mathcal{K}}_\mathbf{k}\times \frac{d\pmb{\mathcal{K}}_\mathbf{k}}{dt}.
\end{equation}
The solutions to these equations in the presence and absence of damping will be presented in the next section.

\section{Results}\label{sec:4}
\begin{figure}
\includegraphics[width=9cm]{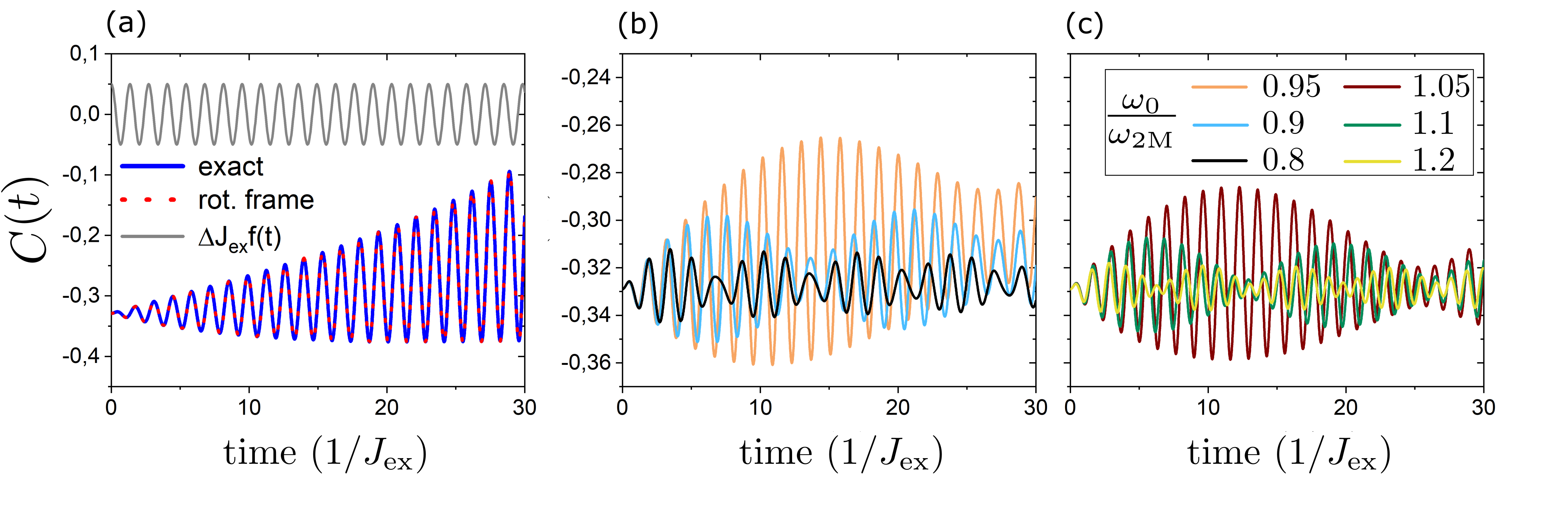}%
\caption{Dynamics of the nearest neighbour correlation $C(t)$ in a $30\times 30$ system with $\Delta J_\text{ex}=0.05$ and $\eta = 0$. (a) Comparison between the numerical exact solution of Eqs. \eqref{eq:theta}-\eqref{eq:phi} (blue solid line) and the rotated frame solution (red dots) at the resonant condition $f(t)=\cos\omega_0 t$ (grey solid line) with $\omega_0 = \omega_\text{2M}$. (b)-(c) Correlation dynamics for several driving frequencies respectively below and above the resonant case $\omega_0 = \omega_\text{2M}$. }\label{fig:Fig1}
\end{figure}

The possibility of exciting parametric resonance of magnon pairs enables to perform logic operations by employing the stable oscillations of the parametrically enhanced magnon modes.
To illustrate this, we focus on the dynamics of spin-spin correlations between two neighbouring spins along the x direction of the square lattice $C(t) = \braket{\hat{S}(\mathbf{r}_i)\cdot \hat{S}(\mathbf{r}_i+\pmb{\delta}_x)}$, which can be accessed experimentally with pump-probe techniques \cite{Zhao2004}. In terms of the pseudo-spin variables, and up to a constant term, they can be expressed as $C(t)=\frac{2S}{N} \sum_\mathbf{k} C_\mathbf{k}(t)$, with
\begin{equation}\label{eq:corr}
C_\mathbf{k}(t) =  \frac{2\mathcal{K}^z_\mathbf{k}\big(1-\gamma_\mathbf{k}\cos(k_x)\big) + \mathcal{K}_\mathbf{k}^x \big( \cos(k_x)-\gamma_\mathbf{k} \big)}{\sqrt{1-\gamma_\mathbf{k}^2}}.
\end{equation}
Note that in $C(t)$ each mode contributes to the dynamics. However, due to a van Hove singularity in the magnon density of states, most of the magnons excited have wave vectors near the edge of the Brillouin zone. This, and the fact that nearest neighbour correlations are mainly determined by high-$\mathbf{k}$ magnons, signifies that driving frequencies close to twice the frequency of zone-edge excitations $\omega_\text{2M}=2\omega_{(\pi,0)}$, result in a strong signal of $C(t)$. In this regime, we can approximate $C(t)\approx B\,C_{(\pi,0)}$ and restrict analytical calculations to $\mathbf{k}=(\pi,0)$. Here, $B$ is a constant proportional to the magnon density of states close to the Brillouin zone.

\begin{figure}
\includegraphics[width=9.cm]{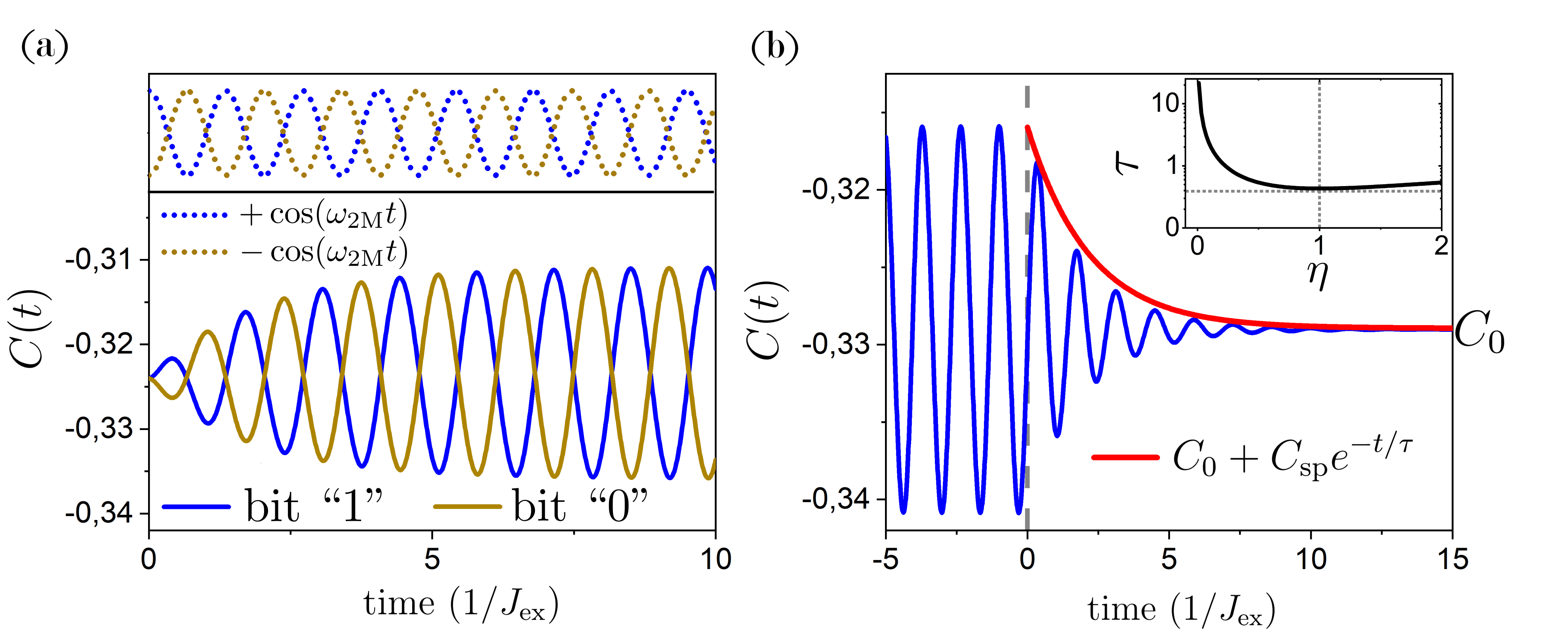}%
\caption{(a) Out of phase dynamics of $C(t)$ triggered by a driving pulse with $f(t)=\cos(\omega_\text{2M}t)$ (blue line) and $f(t)=-\cos(\omega_\text{2M}t)$ (light-brown line) representing respectively the bit "1" and the bit "0".  (b) Decay of $C(t)$ after the pulse is switched off at $t=0$, compared with the decay function $C_0+C_\text{sp}e^{-t/\tau}$, where $C_\text{sp}$ is the amplitude of $C(t)$ at the stable oscillation state and $C_0$ is the value of $C_x(t)$ for $t\rightarrow \infty$. In both panels a system with $30\times 30$ spins is considered, with $\Delta J_\text{ex}=0.05$ and $\eta=0.1$. The inset shows the decay time $\tau$ defined in Eq. \eqref{eq:rel_sol} as a function of the damping parameter $\eta$.}\label{fig:Fig2}
\end{figure}

The dynamics of $C(t)$ obtained by solving Eqs. \eqref{eq:theta}-\eqref{eq:phi} with $\eta = 0$ is shown in Fig. \ref{fig:Fig1} for $\omega_0 = \omega_\text{2M}$ (Fig. \ref{fig:Fig1}(a)) and different driving frequencies below (Fig. \ref{fig:Fig1}(b)) and above the resonant frequency (Fig. \ref{fig:Fig1}(c)). Interestingly, the results show that at the parametric resonance the dynamics in the rotating frame approximation (red dots in Fig. \ref{fig:Fig1}(a)) obtained from Eqs. \eqref{eq:rot_frame_sol} are in excellent agreement with the numerically exact solution of Eqs. \eqref{eq:theta}-\eqref{eq:phi} (without damping).
In the presence of damping and for $\omega_0=\omega_\text{2M}$, correlations
evolve qualitatively similar to the zero damping case, until they reach a stationary condition with oscillations at frequency $\omega_0$ and constant amplitude.
Depending on the sign of the perturbation, it is possible to drive the system into two different states with opposite oscillation phases, as shown in Fig. \ref{fig:Fig2}(a).
The amplitude of such oscillation states can be obtained analytically from the rotating frame approximation of Eqs. \eqref{eq:theta}-\eqref{eq:phi} demanding $\phi_\text{sp}(t)=\omega_\text{2M}t+\phi_\text{sp}(0)$, where $\theta_\text{sp}$ is a constant to be determined \footnote{In the rotating frame approximations (i) and (ii) the equations of motion read: $\dot{\theta}_\mathbf{k} =  V_\mathbf{k}\sin(\phi_\mathbf{k}-\omega_0 t) - \eta\sinh(\theta_\mathbf{k})\dot{\phi}_\mathbf{k}$ and $\dot{\phi}_\mathbf{k} = 2\omega_\mathbf{k} +  V_\mathbf{k} \cos(\phi_\mathbf{k}-\omega_0 t)\coth \theta_\mathbf{k}+\frac{\eta}{\sinh\theta_\mathbf{k}}\dot{\theta}_\mathbf{k}$.}. It is found that
\begin{eqnarray}\label{eq:sp_sol}
\mathcal{K}^x_\text{sp} =-\frac{\mathcal{K} V_\mathbf{k}}{\eta\omega_\text{2M}}\sin(\omega_\text{2M}t), \quad \mathcal{K}^z_\text{sp} = \mathcal{K} \sqrt{1+\frac{V^2_\mathbf{k}}{\eta^2\omega^2_\text{2M}}},
\end{eqnarray}
and $\phi_\text{sp}(0)=-\pi/2$.
Since $\mathcal{K}_\text{sp}^z$ is constant, the amplitude of the correlations at the stationary oscillation states is proportional to $\mathcal{K}_\text{sp}^x$ and thus to $1/\eta$. If instead the system is at one of the two oscillation states and the driving pulse is switched off, the correlations will relax due to the presence of damping. For this case, it is possible to analytically solve the equations of motion, obtaining $\phi_\mathbf{k}(t) = {\omega_\text{2M}t}/(1+\eta^2)-\pi/2$ and $\theta_\mathbf{k}(t)$ such that
\begin{equation}\label{eq:rel_sol}
    \tanh\Big(\frac{\theta_\mathbf{k}}{2}\Big) = \tanh (\theta_\text{sp})\,e^{-\frac{t}{\tau}},\quad \text{with}\quad \tau = \frac{1+\eta^2}{\eta \omega_\text{2M}}.
\end{equation}
Hence, the correlations relax to zero with a relaxation time $\tau$. The dynamics of $C(t)$ obtained numerically solving Eq. \eqref{eq:theta}-\eqref{eq:phi} during the relaxation stage is plotted in Fig. \ref{fig:Fig2}(b) for $\eta=0.1$ (blue line), showing that the envelope of the oscillations indeed decays as $e^{-t/\tau}$ (red line).

\begin{figure}
\includegraphics[width=10cm]{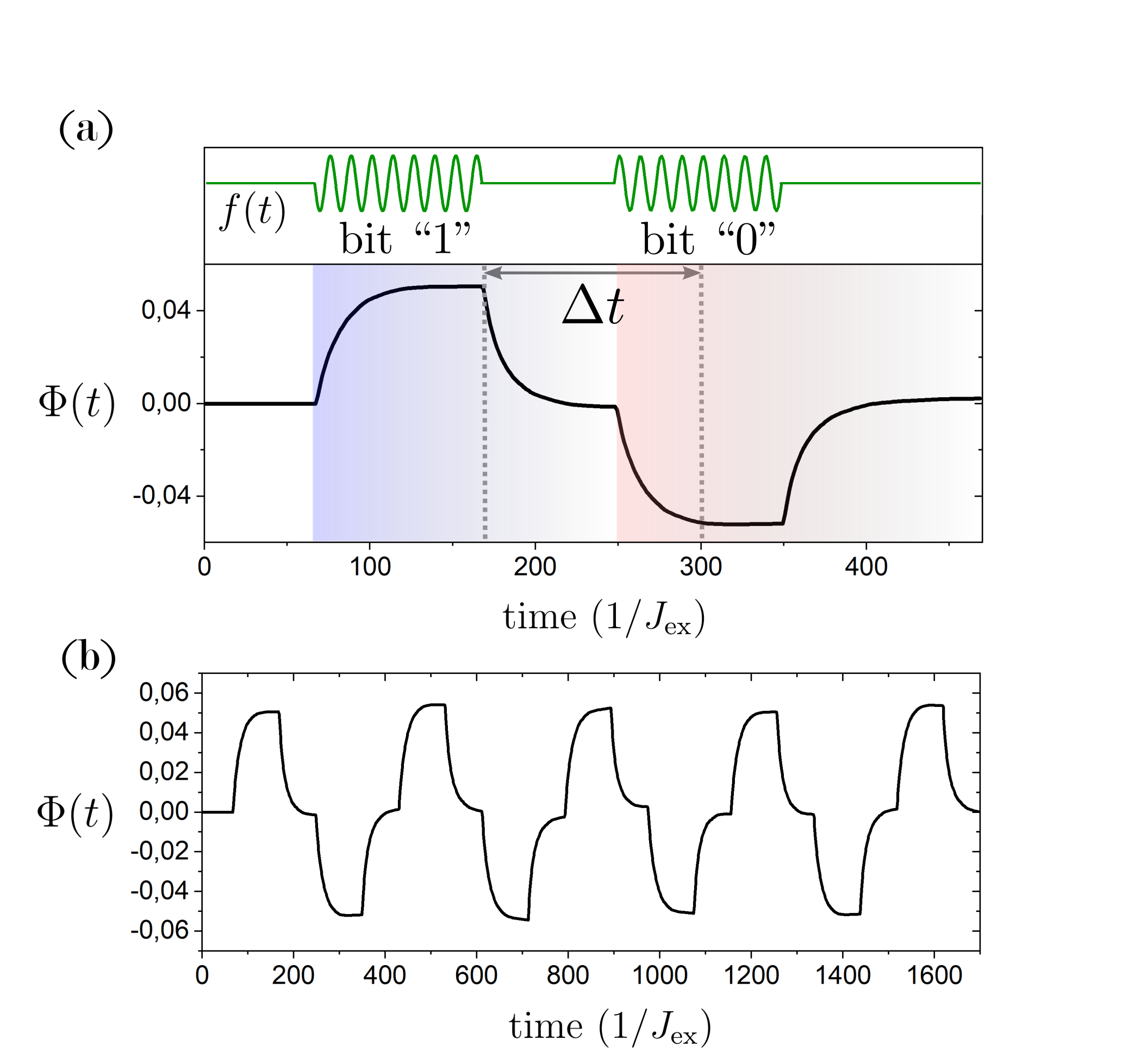}%
\caption{(a) Single switching triggered by two pulses with opposite phase. (b) Multiple switching events achieved by alternating driving pulses with opposite phase. Both are obtained for a system with $30\times 30$ spins,  $\Delta J_\text{ex}=0.05$ and $\eta=0.01$.}\label{fig:Figswitch}
\end{figure}

Employing parametric oscillations of magnon-pairs to perform logic operations requires, minimally, that a switching protocol can be performed between two distinguishable states. In a classical parametron, these states are represented by the two stable phases of the parametric oscillator \cite{Goto1959}. A switching protocol can then be defined for the magnon case as the evolution between the two opposite phases of the oscillations, which can be indicated as "1" and "0". This can be realized using a sequence of two pulses: the first pulse excites the parametric resonance driving the system in either of the two phases of the oscillations corresponding to the bits "1" or "0"; as long as the pulse is active the system will remain in one of these two oscillation states. If the pulse is switched off, due to the presence of damping the amplitude of the oscillations will relax exponentially to zero. When the oscillations are sufficiently small the second pulse with opposite sign is switched on and induces oscillations with opposite phase. In order to single-out the two oscillation states, the crucial information that we need to extract is the phase of $C(t)$. This can be read out by a lock-in amplifier with a reference signal oscillating at the same phase of either one of the two stable states of the oscillations. By choosing as a reference signal $f_{\text{ref}}(t)=\cos(\omega_\text{2M}t)$, the phase $\Phi(t)$ can be obtained from $\Phi(t) = \int_t^{t+\frac{2\pi}{\omega_\text{2M}}}C(t')\cdot f_{\text{ref}}(t') dt'$.
A single switching event between the state "1" and "0" is shown in Fig. \ref{fig:Figswitch}(a), where the dynamics of the correlations is calculated by solving Eqs. \eqref{eq:theta}-\eqref{eq:phi} with $\eta = 0.01$ and $\Delta J_{\text{ex}}/J_{\text{ex}}= 0.05$ in a $30\times 30$ lattice. Importantly, in the stationary oscillation state the magnon decay is exactly compensated by the driving field. Therefore, the amplitude of the signal does not degrade with time, which allows to repeat the switching protocol with no (theoretical) limitations, as shown in Fig. \ref{fig:Figswitch}(b).

Two main parameters are now investigated in relation with the speed and energy efficiency of the proposed magnon-pair parametron: the switching time $\Delta t$ and the switching energy $\Delta E$.  As indicated in Fig. \ref{fig:Figswitch}(a), the switching time is determined by the time taken by $C(t)$ to relax to zero and by the time required to build-up the oscillations to the stationary state.
To calculate the energy dissipated, we note that during a phase-reversal event power dissipation can be evaluated from the work done by the Rayleigh dissipation function. This is given by $P(t) = 2 {F}[\phi,\theta,\dot{\phi},\dot{\theta}]$, and the energy dissipated during a switching event can be obtained from $ \Delta E = \int_{t_0}^{t_0+\Delta t} P(t')dt'$, where $t_0$ is the time at which the pulse is switched off.
Both $\Delta E$ and $\Delta t$ can be computed numerically from the solutions of Eqs. \eqref{eq:theta}-\eqref{eq:phi}, but to gain further insights into the role of $\eta$, approximate analytical expressions are more useful. To this purpose, we assume that the total $\Delta E$ and $\Delta t$ are given by twice the switching energy and time of the relaxation stage, yielding a lower bound for $\Delta E$ and $\Delta t$.  During the relaxation, correlations decay with relaxation time  $\tau$. Taking $\tau$ as the time after which $90\%$ of the amplitude of $C(t)$ is reduced as compared to its value in the stable oscillation state, we obtain a decay time of $\approx 2\tau$ and therefore $\Delta t \approx 4 \tau$. Given the $\eta$-dependence of $\tau$, we get that $\Delta t$ decreases with increasing $\eta$ for $\eta \leq 1$, but it increases when $\eta>1$, as shown in the inset of Fig. \ref{fig:Fig2}(b). The energy is obtained by integrating the Rayleigh dissipation function for $\mathbf{k}=(\pi,0)$ during the relaxation regime and by multiplying the outcome by the number of modes close to the Brillouin edge $B$. This yields 
\begin{equation}
    \Delta E\approx 2B\mathcal{K}\bigg(  \sqrt{\omega_\text{2M}^2 + \frac{V^2_\mathbf{k}}{\eta^2}} -\omega_\text{2M}\bigg)
\end{equation}
It is straightforward to see that $\Delta E$ decreases as $\eta$ increases, tending to zero at large $\eta$.
In the next section, we compare this result with the well-known physical limits of energy and time for computing and quantitatively with existing results in the literature.

\begin{figure}
\includegraphics[width=9cm]{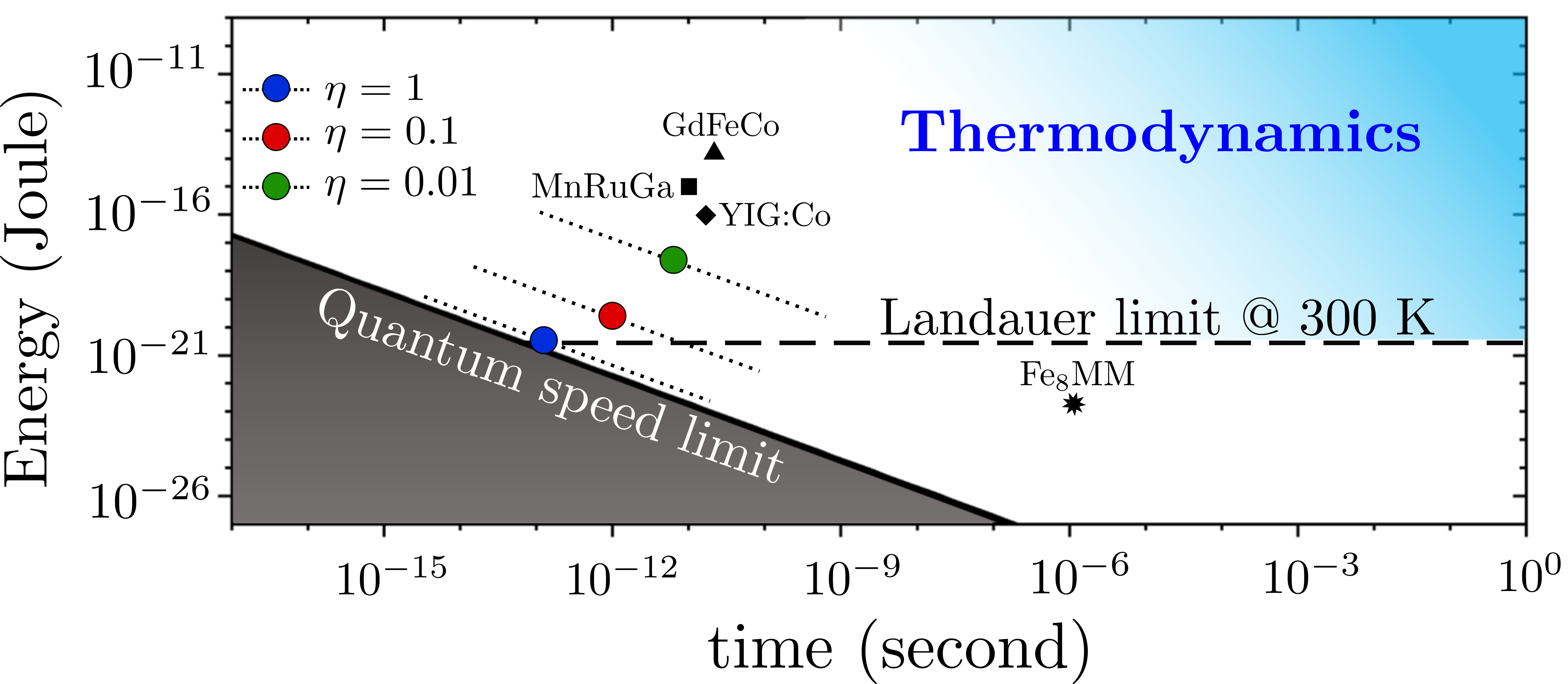}%
\caption{Energy-time diagram showing the performance of the magnon-based parametron for different damping $\eta$ for $J_\text{ex}=0.01$ eV, with the dotted lines moving along the constant $\Delta E \Delta t $ as a function of $J_\text{ex}$. These are compared with the best known experimental results for switching in GdFeCo \cite{Stanciu2007, Vahaplar2009} (triangle), YIG:Co \cite{Stupakiewicz2017} (diamond) and in MnRuGa \cite{Banarjee2020, Banerjee2021} (square) when scaled to a bit of dimension $40\text{ nm}\times 40\text{ nm}\times 20$ nm and  with the quantum switching in the nanomagnet Fe$_8$MM \cite{Gaudenzi2017} (star). The figure also reports the quantum speed limit $\Delta E \Delta t = 1.66 \cdot 10^{-34}$ Joule$\cdot$s (black solid line) and the Landauer limit at 300 K (black dashed line). }\label{fig:QSL}
\end{figure}

\section{Discussion}\label{sec:5}
Determining the physical limits for the time and energy of binary computations has gained considerable interest in recent years, both for technological reasons and for fundamental research at the crossroads between information theory, thermodynamics and condensed matter. Exploiting the laws of thermodynamics,  Landauer \cite{Landauer1961} showed that an irreversible binary operation has a minimal energy expense of $\Delta E_L=(\ln 2)k_B T$, where $k_B$ is the Boltzmann constant and $T$ is the absolute temperature. At room temperature, the Landauer principle (LP) yields an energy dissipation of $\approx 3\cdot 10^{-21}$ J. This limit was verified experimentally using (semi-)classical systems, which demonstrated that the Landauer limit can only be reached in the long-time limit. In particular, long switching times of $\sim 10^{-3}$--$10^{1}$ seconds are needed to minimize unnecessary heat production \cite{Berut2012, Jun2014,Madami2014, Hong2016}. 

Nevertheless, this can dramatically change beyond the realm of irreversible computing. In fact, since the early days of quantum mechanics it is known that energy and time are intrinsically related as it appears from the uncertainty relation $\Delta E \Delta t \geq \hbar$, which can be inferred from the wave-like property of quantum mechanics. While the interpretation of the latter is often debated, the link between $\Delta E $ and $\Delta t$ can be put on a firmer ground with the concept of quantum speed limit (QSL) \cite{Deffner2017}. Roughly speaking, a QSL relates the time it takes to evolve between two orthogonal quantum states with the energy difference $\Delta E$ between these two states, yielding:
\begin{equation}
\Delta E \Delta t \geq h/4 \sim 1.66\cdot 10^{-34} \text{ Joule}\cdot\text{s}.
\end{equation}
For a magnetic system, this limit can be understood semi-classically and the equality sign is a manifestation of precessional switching \cite{Gerrits2002} for a single spin \cite{Reijmer2019}. A recent work demonstrated that it is indeed possible to enhance the switching speed at the Landauer's limit by employing a quantum set-up \cite{Gaudenzi2017}. However, the results obtained in this work are still five orders of magnitude away from the QSL. 

In addition, despite that the exact quantum evolution is reversible and hence in principle can occur without any heat dissipation in the physical computing system, it remains an open question if it is in principle possible to reach the QSL while dissipating at most the energy cost of $\Delta E_L$ at room temperature ($T=300$K). This "ultimate" switching event would require a switching time of no more than $\Delta t = h/(4\Delta E_L)\sim 60$ fs, as shown in Fig. \ref{fig:QSL}. 

For the protocol proposed here and by combining the switching time and energy found in the previous section, we obtain that, as a function of $\eta$, $\Delta E \cdot \Delta t$ has its minimum at $\eta=1$, for which, by numerically solving Eqs. \eqref{eq:theta}-\eqref{eq:phi} for $\Delta J_\text{ex}/J_\text{ex}=0.05$,  we get that $\Delta E \cdot \Delta t \sim 2.5\, h/4$.
This result is plotted in Fig. \ref{fig:QSL} for $J_\text{ex}=10^{-2}$ eV (blue dot), together with the corresponding values for $\eta = 0.1$ ($\Delta E \cdot \Delta t \sim 10^2\, h/4$, red dot) and $\eta= 0.01$ ($\Delta E \cdot \Delta t \sim 5\cdot 10^4\, h/4$, green dot). The dotted lines in Fig. \ref{fig:QSL} instead span all the realistic values of $J_\text{ex}$ along the line of constant $\Delta E \cdot\Delta t$ for each $\eta$.

This result shows  that for the optimal value of the damping, the switching protocol proposed here can border the quantum speed limit. However, it is important to take into account that at $\eta = 1$ the correlations amplitude is strongly quenched with respect to the undamped case at $\eta = 0$. In fact, appreciable (parametric) amplifications of the correlations signal can only be observed for $\eta \lesssim 0.1$ and thus two orders of magnitude away from the quantum speed limit. Nevertheless, this is still three orders of magnitude closer to the QSL than the fastest and least dissipative magnetic switching event demonstrated experimentally so far. To conclude this section, we note that in the switching protocol here adopted, the system is initialized in one oscillation state and it is driven to the opposite state. This differs from the erasure Landauer protocol \cite{Landauer1961}, which is a thermodynamical process and is realized between two stable phases of a material. Moreover, in the erasure protocol, the final state is fixed and the initial state can be in either of the two possible states, while  here the initial and final oscillation states are both fixed. Nevertheless, an analogous erasure process can be realized in the magnon-pair parametron and leads to the same switching time and energy found for the protocol considered in Fig. \ref{fig:Figswitch}. The switching time and energy, indeed, only depend on $\eta$ and the amplitude of the correlations at the stable oscillation state, which, in turn, only depends on the physical parameters of the system and not on how the switching protocol is performed.

\section{Conclusions}\label{sec:6}
In this letter we have shown that periodic modulations of the exchange interactions can lead to parametric oscillations of short wavelength, THz frequency magnon pairs. Following the analogy with a classical parametron architecture, we developed a semi-classical approach to magnon pair dynamics and showed that bit reversal can be realized by switching between two opposite phases of the parametric oscillations. In addition, we predicted that the switching time and energy dissipated approaches the quantum speed limit for switching between two oscillation states 5 orders of magnitude closer than is experimentally demonstrated so far, with an energy dissipation that touches the thermodynamic bound set by the Landauer limit. To the best of our knowledge, this is the fastest and least dissipative switching protocol predicted for magnetic materials so far and we anticipate that this work may stimulate further experimental research on ultrafast antiferromagnetic switching driven by THz pulses.

Further theoretical work may focus on understanding how to combine two or more magnon-pair parametric oscillators. In principle this allows to perform other elementary logic operations such as AND, OR and XOR\cite{Goto1959}. Owing to the small wavelength of the magnons excited, the logic unit may be engineered to be as small as few nanometers, and at the same time, several orders of magnitude faster and more energy-efficient than the current technology based on CMOS transistors. Moreover, by using parametric excitation at different frequencies, several parametric oscillators may be constructed within one material.

The current theoretical model employed shows that the switching time and energy both decrease by increasing the damping field. The switching protocol seems therefore most suitable for metallic antiferromagnets, where typically a strong magnon damping is found\cite{Liu2017,Mahfouzi2018,Simensen2020}. On the other hand, when the damping is weak, magnon-magnon interactions themselves will have a significant role on the damping of magnon-pair oscillations \cite{Kotthaus1972, Lockwood1973, Bayrakci2006, Bayrakci2013}, opening new avenues for all-coherent reversal between different oscillation modes at the quantum speed limit. 

Finally, we note that the switching protocol here introduced differs from conventional scenarios where the switching is performed between two stable magnetization domains. Due to the analogy between the pseudo-spin dynamics of the SU(1,1) algebra and the classical magnetization precession in the SU(2) sphere, we expect that an analogous precessional switching can be envisaged for the pseudo-spin vector $\pmb{\mathcal{K}}_\mathbf{k}$. In addition, the simplified description on the SU(1,1) unit sphere might help to disclose new coherent switching protocols, in which strongly nonlinear dynamics of magnon-pairs is employed to switch between different stationary antiferromagnetic states at the fastest time scale and with the least amount of energy dissipation.

\begin{acknowledgments}
We acknowledge stimulating discussions with A.V. Kimel. This work is part of the Shell-NWO/FOM-initiative ``Computational sciences for energy research” of Shell and Chemical Sciences, Earth and Life Sciences, Physical Sciences, FOM and STW, and received funding from the European Research Council ERC grant agreement No. 856538 (3D-MAGiC). 
\end{acknowledgments}

\bibliography{references}

\end{document}